# Additive engineering for $Sb_2S_3$ indoor photovoltaics with efficiency exceeding 17%

Xiao Chen,[1] Xiaoxuan Shu,[2] Jiangcheng Zhou,[1] Lei Wan,[1] Peng Xiao,[3] Yuchen Fu,[4] Junzhi Ye,[4] Yi-Teng Huang,[4] Bin Yan,[5] Dingjiang Xue,[5] Tao Chen,[3] Jiejie Chen*,[2] Robert L. Z. Hoye,*,[4] and Ru Zhou*,[1,4]

[1] School of Electrical Engineering and Automation, Hefei University of Technology, Hefei 230009, P. R. China

[2] Key Laboratory of Urban Pollutant Conversion, Department of Environmental Science and Engineering, University of Science & Technology of China, Hefei 230009, P. R. China

[3] Hefei National Research Center for Physical Sciences at the Microscale, School of Chemistry and Materials Science, University of Science and Technology of China, Hefei 230026, P. R. China

[4] Inorganic Chemistry Laboratory, Department of Chemistry, University of Oxford, South Parks Road, Oxford OX1 3QR, United Kingdom

[5] Beijing National Laboratory for Molecular Sciences, CAS Key Laboratory of Molecular Nanostructure and Nanotechnology, Institute of Chemistry, Chinese Academy of Sciences, Beijing 100190, P. R. China

* Corresponding authors:
chenjiej@ustc.edu.cn (Jiejie Chen)
robert.hoye@chem.ox.ac.uk (Robert L. Z. Hoye)
zhouru@hfut.edu.cn (Ru Zhou)

# Abstract

Indoor photovoltaics (IPVs) have attracted increasing attention for sustainably powering Internet of Things (IoT) electronics. $Sb_2S_3$ is a promising IPV candidate material with a bandgap of ~1.75 eV, which is near the optimal value for indoor energy harvesting. However, the performance of $Sb_2S_3$ solar cells is limited by nonradiative recombination, closely associated with the poor-quality absorber films. Additive engineering is an effective strategy to improved the properties of solution-processed films. This work shows that the addition of monoethanolamine (MEA) into the precursor solution allows the nucleation and growth of $Sb_2S_3$ films to be controlled, enabling the deposition of high-quality $Sb_2S_3$ absorbers with reduced grain boundary density, optimized band positions and increased carrier concentration. Complemented with computations, it is revealed that the incorporation of MEA leads to a more efficient and energetically favorable deposition for enhanced heterogeneous nucleation on the substrate, which increases the grain size and accelerates the deposition rate of $Sb_2S_3$ films. Due to suppressed carrier recombination and improved charge-carrier transport in $Sb_2S_3$ absorber films, the MEA-modulated $Sb_2S_3$ solar cell yields a power conversion efficiency (PCE) of 7.22% under AM1.5G illumination, and an IPV PCE of 17.55% under 1000 lux white light emitting diode (WLED) illumination, which is the highest yet reported for $Sb_2S_3$ IPVs. Furthermore, we construct high performance large-area $Sb_2S_3$ IPV modules to power IoT wireless sensors, and realize the long-term continuous recording of environmental parameters under WLED illumination in an office. This work highlights the great prospect of $Sb_2S_3$ photovoltaics for indoor energy harvesting.

# 1. Introduction

The rapid development of Internet of Things (IoT) technologies is leading to an ongoing exponentially growing market of smart devices[1,2]. Such end nodes are usually designed for cyclic operation and low power consumption (microwatts to milliwatts), and having a reliable and sustainable long-term power supply is critical for the success of IoT technologies. Currently, autonomous IoT nodes are mostly powered using batteries. However, the short lifespan of batteries not only limits the power consumption and size of IoT devices, but also restricts the applications to the cases which are compatible with battery replacement and maintenance. Solely relying on batteries to power IoT devices might not sustain the rapidly growing size of the IoT ecosystem as it proceeds to a trillion nodes, where it is predicted that >100 billion batteries will have to be disposed of each year[3]. In this regard, integrated energy systems for harvesting ambient energy (*e.g.*, indoor lighting, mechanical energy, or thermal energy) are now expected to serve as alternatives, or work complementarily to batteries[1,4,5]. Indoor photovoltaics (IPVs), which capture energy from ambient lighting (either from artificial light sources, or daylight), have significant potential to provide sustainable power for driving wireless IoT nodes that communicate using a range of protocols, such as Bluetooth low energy, RFID tags, LoRa, passive Wi-Fi, Zigbee, ANT, etc[6,7]. Indeed, IPVs are deployable in view of their reliance on radiative energy transfer and indoor lighting is ubiquitously available and predictable. Furthermore, IPVs afford relatively high energy density compared to other ambient energy harvesting technologies[3]. Therefore, the development of high performance IPVs is important for sustainable IoT applications[8]. Indoor illuminances typically range from 200-500 lux for private areas, and 500-1500 lux for office, commercial and industrial areas. For commonly used cold white light emitting diodes (WLEDs), an illuminance of 500-1000 lux (0.14-0.28 mW cm$^{-2}$) is 300-700 times lower compared to "1-sun" (AM1.5G, 100 mW cm$^{-2}$)[9]. The narrow emission spectra of indoor light sources (*e.g.*, LEDs and fluorescent lamps (FLs)) range from 400 to 700 nm, which determines the optimal bandgap for indoor light-absorbing materials to be around 1.80-2.00 eV[10]. Therefore, crystalline silicon (c-Si), which dominates the outdoor PVs market, would not be well suited for indoor energy harvesting, due to its small bandgap (~1.12 eV), as well as high dark current densities, which together limit its performance under the lower irradiances from indoor lighting[11]. The state-of-the-art commercial solution up to now for IPVs is hydrogenated amorphous silicon (a-Si:H), with a bandgap of 1.7-2.0 eV. The PCEs of commercially available standard a-Si:H module devices typically range from 4.4%-9.2%[3]. To date, emerging next-generation solar cells, including organic photovoltaics (OPVs), dye-sensitized solar cells (DSSCs) and lead-halide perovskite solar cells (LHPSCs) have also been investigated for IPV applications[8,12-16]. However, DSSCs and OPVs are limited by the use of toxic solvents and expensive small molecules; LHPSCs involve the use of high contents of lead, which exceeds the limits of the EU restriction of hazardous substances directive (RoHS), and this may act as a barrier to commercialization. A stable material that can be simply processed and is comprised of RoHS-compliant elements is missing.

$Sb_2S_3$ is an earth-abundant, low-toxicity and stable material with a bandgap of ~1.75 eV, close to the optimal bandgap value for IPV application[10]. Moreover, $Sb_2S_3$ possesses excellent materials and optoelectronic properties, including simple binary composition, quasi-one-dimensional crystal structure, high absorption coefficients ($10^4 \sim 10^5$ cm$^{-1}$), low melting points (~500 °C) and high vapor pressures. These merits enable the low-temperature fabrication of high performance, flexible and lightweight $Sb_2S_3$ devices for powering IoT electronics. According to calculations by Hoye et al., the spectroscopic limited maximum efficiency (SLME) reaches 47% under 1000 lux white LED lighting[3]. However, up to now, the development of $Sb_2S_3$ solar cells has mainly focused on its performance under 1-sun illumination (*i.e.*, outdoor photovoltaics). Since the illumination intensity and emission spectra of indoor light sources in most building and office environments, such as FLs and LEDs, are strikingly different from 1-sun conditions, it is necessary to carefully study the IPV performance of $Sb_2S_3$ photovoltaics[4,17]. As far as we know, only Zheng et al. reported the photovoltaic performance under low-intensity illuminations for $Sb_2S_3$ devices, which delivered 16.37% indoor efficiency illuminated by a 1000 lux WLED[18]. The application of $Sb_2S_3$ devices for powering IoT electronics has, to our knowledge, never been demonstrated.

Thanks to endeavors from a growing community of researchers focusing on this chalcogenide photovoltaic materials system, great achievements have been made for $Sb_2S_3$ solar cells. To date, the mesoporous and planar $Sb_2S_3$ photovoltaics have delivered record efficiencies of 7.5% and 8.0% under 1-sun illumination, respectively[19,20]. However, such efficiency values still lag far behind the theoretical maximum efficiency of 28.64% according to the Shockley-Queisser limit[21]. The severe charge-carrier recombination at grain boundaries (GBs) and interfaces is one of the critical problems that limits the performance of $Sb_2S_3$ solar cells[21,22]. Over the past decade, researchers have explored a variety of approaches to enhance the efficiency of $Sb_2S_3$ solar cells, such as ion doping[21,23], additive engineering[18,24], surface passivation[25], configuration optimization[26], etc. Additive engineering has attracted tremendous attention because it is a simple strategy to regulate the deposition processes as well as the film properties for solution-processed $Sb_2S_3$ films, such as deposition rate, growth orientation, film morphology, etc[20]. To date, a number of additives, including sodium dodecyl sulfate[18], thioacetamide[20], tartaric acid[27], tetramethylammonium hexafluoro phosphonate[24], and ammonium sulfide[28] have been investigated to adjust the chemical bath deposition or hydrothermal deposition processes of $Sb_2S_3$ films. For instance, Wang et al. used thioacetamide together with sodium thiosulfate to act as dual sulfur sources and achieved noticeable improvements in the device efficiency due to the advantageous hydrolysis cooperation between two sulfur sources[20]. Han et al. employed tetramethylammonium hexafluoro phosphonate which could coordinate with $Sb^{3+}$ due to the multidentate anchoring and further improve the morphology of $Sb_2S_3$ films as well as reduce the trap states[24]. Therefore, additive engineering is promising to optimize the film properties and further enhance the photovoltaic performance of $Sb_2S_3$ solar cells.

As a complexing agent that has hydrophilic hydroxyl group (-OH) and amino group ($-NH_2$), monoethanolamine (MEA) is expected to be capable of effectively regulating the nucleation and growth environment for film deposition[29,30]. In this work, we obtained high-quality $Sb_2S_3$ films through an additive-modulated hydrothermal deposition method by incorporating a small amount of MEA into the precursor solution. This additive engineering strategy increases the grain size and accelerates the deposition rate of $Sb_2S_3$ films, which further improve the optoelectronic properties of absorber films. The MEA-modulated planar $Sb_2S_3$ solar cells delivered an impressive PCE of 7.22% under 1-sun illumination and a remarkable indoor PCE of 17.55% under 1000 lux LED illumination. We further demonstrated the use of large-area $Sb_2S_3$ modules to power low-power portable electronics.

## 2. Results and discussion

### 2.1. Deposition of high-quality $Sb_2S_3$ films

In this work, we employed a hydrothermal deposition method to deposit $Sb_2S_3$ films onto $FTO/SnO_2/CdS$ substrate. Antimony potassium tartrate (APT) and sodium thiosulfate were used as antimony source and sulfur source, respectively. A small amount of MEA was used as an additive in the precursor solution to regulate the deposition of $Sb_2S_3$ films. Fig. 1a shows top-view scanning electron microscopy (SEM) images of as-deposited $Sb_2S_3$ thin films prepared without the additive (the control sample), and with different concentrations of MEA additives. For the convenience of description, the $Sb_2S_3$ films obtained with MEA concentrations of 3 μg/mL, 4 μg/mL, and 5 μg/mL in the hydrothermal deposition are labeled as MEA-3, MEA-4, and MEA-5, respectively. As shown, the control sample displays grain sizes smaller than 5 μm, consistent with that reported for hydrothermally-deposited $Sb_2S_3$ films[21]. It is worth noting that we have previously established that SEM images are suitable for evaluating the grain size of $Sb_2S_3$ films[21]. With the addition of MEA, the grain size of $Sb_2S_3$ films involves a significant increase, and some $Sb_2S_3$ grains exceed 10 μm with an MEA concentration of 4 μg/mL. The increase in grain size is accompanied by a decrease in GB density, which can be defined as the GB length per unit area. As shown in Fig. 1c, the GB density of $Sb_2S_3$ films decreases from $0.434 \pm 0.027$ μm cm$^{-2}$ for the control sample to $0.319 \pm 0.006$, $0.303 \pm 0.010$, and $0.229 \pm 0.006$ μm cm$^{-2}$ for the MEA-3,

MEA-4, and MEA-5 samples, respectively. The reduced GB density of the absorber films is expected to favor the suppression of non-radiative recombination[31]. The cross-sectional SEM images (Fig. 1b) reveal that the $Sb_2S_3$ films shows a gradual increase of thickness with the addition of MEA. The film thickness increases from 328 ± 11 nm for the control sample to 399 ± 8 nm, 428 ± 6 nm, and 480 ± 5 nm for the MEA-3, MEA-4, and MEA-5 samples, respectively. This implies that the addition of MEA to the precursor solution accelerates the deposition rate of $Sb_2S_3$ films. The 2D and 3D morphology spatial maps of atomic force microscopy (AFM) characterization (Supplementary Fig. 1) further indicate that the MEA-4 film samples show larger grain sizes with fewer GBs compared to the control samples. Moreover, the addition of MEA also leads to a reduction in the root-mean-square (RMS) roughness of absorber films from 18.4 nm (the control sample) to 16.2 nm (the MEA-4 sample). The flatter $Sb_2S_3$ films should be beneficial for the formation of high-quality heterojunction between the absorber layer and the solution-processed hole transport layer (HTL) on top.

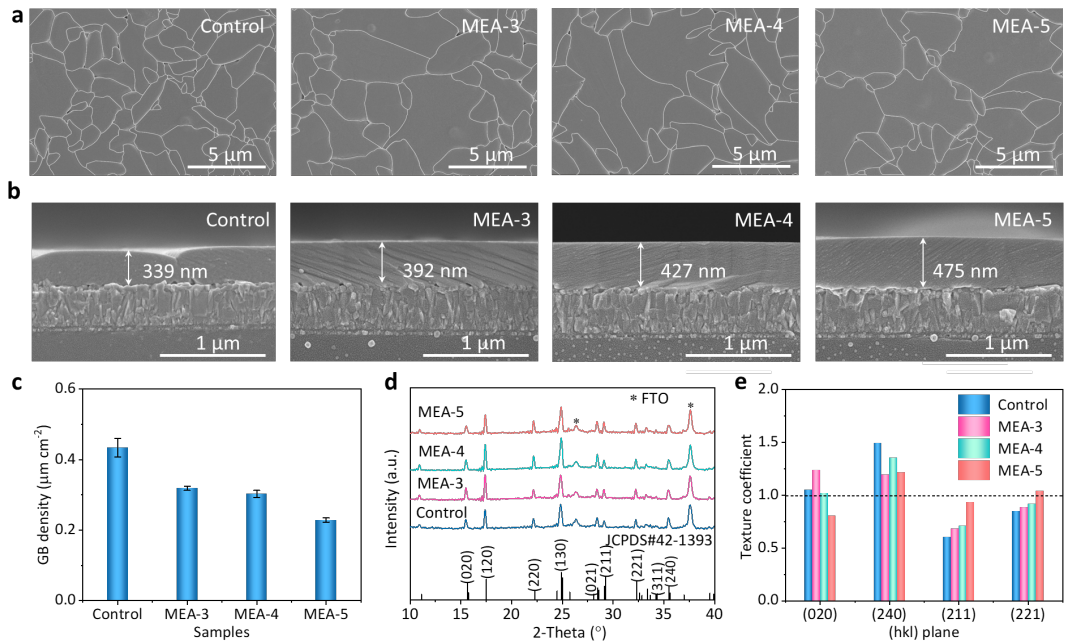

**Fig. 1 Morphological and structural properties of $Sb_2S_3$ films. a, b** Top-view and cross-sectional SEM images and **c** GB densities of hydrothermally deposited $Sb_2S_3$ thin films prepared without (control) and with MEA additives. Three samples were measured to determine the mean GB density values shown, and the error bars represent the standard deviation. **d, e** XRD patterns and texture coefficients of the dominant (020), (240), (211) and (221) peaks of the control film and MEA-$Sb_2S_3$

films prepared with different concentrations of MEA additives.

X-ray diffraction (XRD) patterns of $Sb_2S_3$ films (Fig. 1d) show that the diffraction peaks can be indexed to pure orthorhombic $Sb_2S_3$ (JCPDS #42-1393), with the background peaks from FTO substrates as marked by the asterisks[22]. The diffraction peaks of $Sb_2S_3$ films reveal negligible shifts with the addition of MEA, indicating that the MEA additive could not be incorporated into the $Sb_2S_3$ lattice. Compared to the control film, the MEA-$Sb_2S_3$ films exhibit slightly reduced intensities for (020) and (240) peaks as well as enhanced intensities for (211) and (221) peaks. The texture coefficients (TC) of (020), (240), (211) and (221) planes of $Sb_2S_3$ films are further calculated to evaluate the changes in the film orientation, as shown in Fig. 1e. It can be seen that the addition of MEA results in the (211) and (221) preferred orientations for $Sb_2S_3$ films. As we know, $Sb_2S_3$ has a quasi-1D crystal structure consisting of numerous parallel $(Sb_4S_6)_n$ ribbons with van der Waals interactions between them[22]. Hence the carrier transport along $(Sb_4S_6)_n$ ribbons is much more efficient than the charge hopping between ribbons, and the enhanced (hk1) preferred orientation should favor for the charge-carrier transport in the vertically-structured devices.

Conductive AFM (c-AFM) characterization of the control and MEA-4 films (Supplementary Fig. 2a and b) reveal that the MEA-4 samples display higher current on the film surface than the control sample. The surface current intensity distributions (Supplementary Fig. 2c and d) show that both samples have obvious current fluctuations at GBs, which might be caused by the reduction in the carrier density at GBs[21]. Compared to the control sample, the MEA-4 films show reduced current fluctuations and increased average current intensity. This should be associated with the decrease of GBs for the MEA-4 sample. The alleviated current fluctuation on the absorber films would facilitate the collection of charge-carriers. To further evaluate the conductivity change of $Sb_2S_3$ films, we prepared devices with the simple device structure of FTO/$Sb_2S_3$/Au and measured *I-V* curves in the dark (Supplementary Fig. 3). The conductivity of films can be calculated by the equation[26]: $I = \sigma_0 A L^{-1} V$, where $\sigma_0$ is the conductivity, *A* represents the area of the apparatus (0.06 cm$^2$), *L* is the thickness of Sb$_2$S$_3$ layer (here the thicknesses of the control and MEA-4 films are 339 nm and 427 nm, respectively), and *I* and *V* are the current and the voltage, respectively. The estimated conductivities of the control and MEA-4 films are 3.55 × 10$^{-5}$ and 5.96 × 10$^{-5}$ S cm$^{-1}$, respectively. The increase in the conductivity of MEA-4 film might be associated to the increase in grain size, consistent with the c-AFM results[32].

## 2.2. Mechanisms of additive engineering

We now investigate the underlying mechanism for the impact of the MEA additives on the film properties of Sb$_2$S$_3$. First, we hypothesize that the MEA additive is absent in as-deposited Sb$_2$S$_3$ films, since the decomposition temperature of MEA molecules (~200 °C) is much lower than the post-annealing temperature for Sb$_2$S$_3$ films (~370 °C)[33]. The X-ray photoelectron spectroscopy (XPS) and Raman characterization of the control and MEA-4 films were performed to test this hypothesis. XPS spectra (Supplementary Fig. 4) and Raman spectra (Supplementary Fig. 5) reveal that both of the control and MEA-4 film samples share nearly identical characteristic peaks, indicating that the incorporation of MEA additives exerts no impact on the elemental composition and chemical states of Sb$_2$S$_3$ films. Hence we speculate that MEA only plays roles in the hydrothermal deposition process of Sb$_2$S$_3$ absorber films.

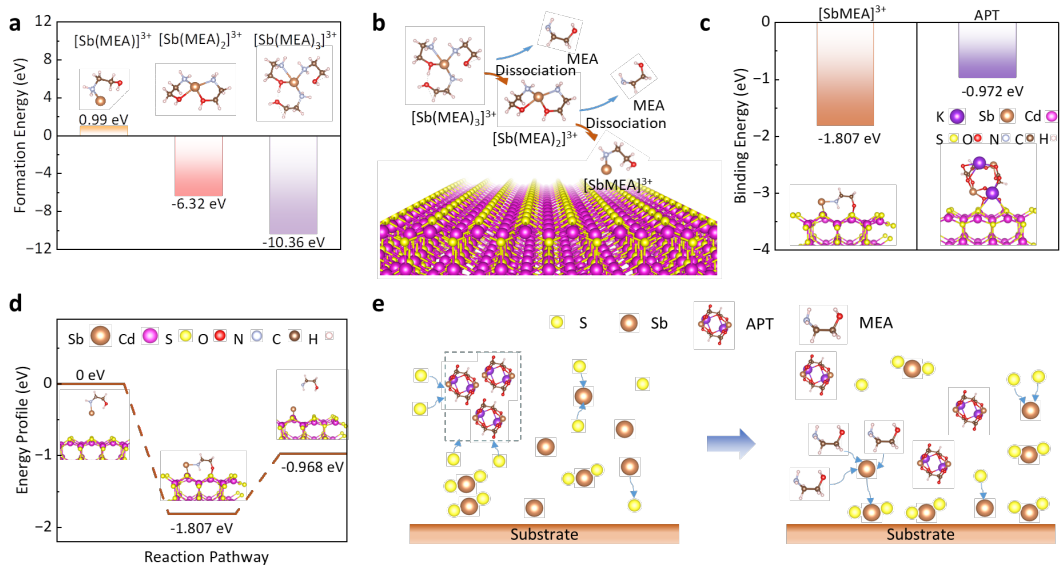

**Fig. 2 Mechanisms of additive engineering for the deposition of Sb$_2$S$_3$ films. a** The formation energy of [Sb(MEA)$_n$]$^{3+}$, *n* = 1, 2, and 3. **b** Schematic diagram depicting the sequential dissociation

of [Sb(MEA)$_3$]$^{3+}$ into [Sb(MEA)$_2$]$^{3+}$ and further into [SbMEA]$^{3+}$ on the CdS substrate with the addition of MEA. **c** The binding energy between the Sb$^{3+}$ complexes ([SbMEA]$^{3+}$ and APT) and the CdS substrate. **d** The reaction energy profile diagram of [SbMEA]$^{3+}$ binding to the CdS substrate and releasing the MEA ligand. **e** Schematic illustration of the evolution of the Sb$_2$S$_3$ deposition process on the substrate with the addition of MEA.

In this study, MEA as the complexing agent was incorporated into the precursor solution to regulate the deposition process. MEA can be uniformly dissolved into the precursor solution. To investigate the possible mechanisms of additive engineering, density functional theory (DFT) calculations were employed to examine the role of ligands and their interactions with the substrate. The models of [Sb(MEA)$_n$]$^{3+}$ complexes were constructed with different coordination numbers, where $n$ = 1, 2, and 3. Coordination numbers exceeding 3 were excluded from consideration due to significant steric hindrance among MEA molecules. As shown in Fig. 2a, [Sb(MEA)$_3$]$^{3+}$ manifests the lowest formation energy, thereby signifying it as the most stable coordination state. Moreover, the dissociation process of [Sb(MEA)$_3$]$^{3+}$ into [Sb(MEA)$_2$]$^{3+}$ and subsequently into [Sb(MEA)]$^{3+}$ necessitates an external energy input of merely 1.003 eV/complex and 0.943 eV/complex, respectively (Fig. 2b and Supplementary Fig. 6). This indicates that continuous thermal energy input into the system allows the coexistence of the three forms of MEA-complexed Sb$^{3+}$ during the hydrothermal deposition stage. Ultimately, Sb$^{3+}$ might bind to the CdS substrate in the form of [SbMEA]$^{3+}$, owing to the complete exposure of Sb$^{3+}$ in this configuration. This specific configuration of the ligand is advantageous for the deposition and growth of Sb$_2$S$_3$ on the substrate. In the hydrothermal deposition of Sb$_2$S$_3$ films, Sb$^{3+}$ ions was introduced in the form of C$_8$H$_4$K$_2$O$_{12}$Sb$_2$, necessitating an examination of the binding capacities of this specific coordination complex at the interface. The strong chelating effect of the tartrate ions introduces complexity to the binding process of Sb$^{3+}$ to the substrate, thereby impeding their effective stabilization on the substrate surface[32]. Moreover, the considerable steric hindrance presented by the molecular structure of C$_8$H$_4$K$_2$O$_{12}$Sb$_2$ further restricts the direct interaction between Sb$^{3+}$ and the substrate. Nevertheless, the incorporation of MEA can markedly mitigate this limitation. The unique bridging capacity of MEA allows its molecular termini to adhere to the substrate, acting as a molecular bridge that facilitates the connection between $Sb^{3+}$ with the CdS substrate. As shown in Fig. 2c, compared to APT, the $[Sb(MEA)]^{3+}$ complex exhibits a more negative binding energy with the CdS substrate. Hence the bridging effect of MEA significantly enhances the affinity and subsequent binding of $Sb^{3+}$ to the CdS substrate, effectively overcoming the problems posed by the chelating effect of tartrate ions and the steric hindrance of APT. During this process, MEA possesses the capability to sequester free $Sb^{3+}$ within the solution, thereby elevating the likelihood of their interaction with the substrate. This dual functionality not only acts as a physical bridge but also enhances the local concentration of $Sb^{3+}$ in proximity to the substrate surface, increasing the deposition rate of $Sb^{3+}$ on the substrate. Furthermore, computational analyses have substantiated that the release of ligands facilitated by the addition of MEA requires minimal energy input, underscoring the high efficiency of MEA in promoting the attachment of $Sb^{3+}$ to the substrate and further the reaction with $S^{2-}$ to form $Sb_2S_3$ (Fig. 2d). Therefore, the MEA additive plays a crucial role in overcoming the challenges faced by APT, culminating in a more efficient and energetically favorable deposition process.

Based on these discussions, the incorporation of MEA into the precursor contributes to the precise control of the deposition conditions, which is essential for the preparation of high-quality $Sb_2S_3$ films[34]. We schematically depict the evolution of the $Sb_2S_3$ deposition process on the substrate with the incorporation of MEA, as shown in Fig. 2e. It is known that, during the initial hydrothermal deposition stage, two typical nucleation processes occur, *i.e.*, homogeneous nucleation in the precursor solution and heterogeneous nucleation on the substrate[21]. In the typical hydrothermal deposition process, the strong chelating effect of ATP would result in the formation of many aggregates in the precursor solutions[32]. These APT aggregates would act as nucleation centers for homogeneous nucleation and growth of $Sb_2S_3$ in the precursor solution. As a result, a large number of suspended $Sb_2S_3$ particles would form in the precursor solution during the hydrothermal deposition without the use of additives[29]. In contrast, the incorporation of MEA results in the formation of MEA-complexed $Sb^{3+}$, which would decrease the amount of APT aggregates and lead to a more uniform distribution of $Sb^{3+}$ in the precursor solution. The formation of fewer aggregates leads to increases in the concentration of $Sb^{3+}$ in the precursor solution, as well as increases in the density of reactive ions on the substrate. The deposition process of $Sb_2S_3$ with the presence of MEA complexing agent can be described by Equations (S1-S8) (Supplementary Note S1)[35]. According to computational analyses, the unique bridging capacity of MEA leads to a decrease in the energy barrier of heterogeneous nucleation and thus promotes the nucleation of $Sb_2S_3$ on the substrate. This further decreases the sites for homogeneous nucleation, in good agreement with the reduction of suspended particles in the reactor at the end of hydrothermal deposition (Supplementary Fig. 7). As a result, the enhanced heterogeneous nucleation and suppressed homogeneous nucleation lead to the formation of dense, uniform and large-grain $Sb_2S_3$ films[21].

Here, to exclude the influence of pH on the deposition rate of $Sb_2S_3$, we measured the pH of the precursor solution without and with the addition of MEA (Supplementary Fig. 8), which reveals that the addition of a small amount of MEA exerts no significant change in the pH of the precursor solution. Therefore, the changes in the morphology and structural properties of $Sb_2S_3$ films should be associated with the addition of MEA. The multifaceted role of MEA plays a pivotal part in regulating the nucleation and growth processes of $Sb_2S_3$ films, resulting in the deposition of ideal absorber films with minimized GB density for constructing solar cells.

### 2.3. Device performance of $Sb_2S_3$ solar cells

We prepared planar $Sb_2S_3$ solar cells with a typical p-i-n configuration of FTO/$SnO_2$/CdS/$Sb_2S_3$/Spiro-OMeTAD/Au, as illustrated in Fig. 3a. The band structures of each functional layers in $Sb_2S_3$ solar cells are shown in Fig. 3b. The energy levels of $Sb_2S_3$ were obtained from ultraviolet photoelectron spectroscopy (UPS) measurements (Supplementary Fig. 9), while the energy levels of other functional layers are compiled from the literature. The detailed analysis of UPS spectra of the control and MEA-4 $Sb_2S_3$ film samples is given in Supplementary Note S2. The corresponding Fermi level ($E_F$), valence band maximum (VBM) and conduction band minimum (CBM) of the control and MEA-4 $Sb_2S_3$ films are summarized in Supplementary Table 1. Here, the bandgap values of the control and MEA-$Sb_2S_3$ films are determined to be 1.74 and 1.73 eV, respectively, based on the Tauc plots of UV-vis-NIR absorption spectra (Supplementary Fig. 10). As shown, the energy levels of $Sb_2S_3$ match well with the electron transport layer (CdS) and HTL (Spiro-OMeTAD), favoring for the efficient charge transport from absorber layer towards two electrodes. Moreover, compared to the control $Sb_2S_3$ film, the Fermi level of MEA-$Sb_2S_3$ is shifted up towards CBM. This indicates an increased electron carrier concentration for n-type $Sb_2S_3$. Compared to the control film, the upward shift of CBM for the MEA-4 film facilitates the photogenerated carrier transport from the absorber layer to the electron transport layer. Fig. 3c gives current density-voltage (*J-V*) curves of best-performing control and MEA-$Sb_2S_3$ solar cells. As shown, the control device delivers a $V_{OC}$ of 766 mV, a $J_{SC}$ of 14.89 mA cm$^{-2}$, an FF of 54.49%, and a PCE of 6.22%, while the MEA-4 device yields an enhanced PCE of 7.22%, with a $V_{OC}$ of 787 mV, a $J_{SC}$ of 16.12 mA cm$^{-2}$, and an FF of 56.92%. That is, the incorporation of MEA affords a 16% relative efficiency enhancement for $Sb_2S_3$ solar cells compared to the control device (or 1% absolute efficiency increase). The performance enhancement of $Sb_2S_3$ solar cells should be associated with the improvements in materials and optoelectronic properties of the $Sb_2S_3$ absorber films. The corresponding external quantum efficiency (EQE) spectra of both devices are shown in Fig. 3d. It can be seen that both devices exhibit a wide light response ranging from 350 to 750 nm. Compared to the control device, the MEA-4 device exhibits an enhanced EQE value close to 90% at around 520 nm. The obtained integrated $J_{SC}$ values are 14.83 and 16.02 mA cm$^{-2}$ for the control and MEA-4 devices, respectively. The values are consistent with the $J_{SC}$ obtained from *J-V* curves (within 5% deviation). Fig. 3e-h displays the distribution statistics of the PCE, $V_{OC}$, $J_{SC}$, and FF for the $Sb_2S_3$ devices, and the mean and champion performance parameters are summarized in Table 1. As shown, the photovoltaic performance of $Sb_2S_3$ solar cells first increases and then decreases with the increasing amount of MEA. MEA-$Sb_2S_3$ devices show enhanced photovoltaic performance compared to the control sample, and an optimal device efficiency can be obtained based on the MEA-4 films. The MEA-4 device delivers an enhanced average efficiency of 6.88% in contrast to that of 5.99% for the control device.

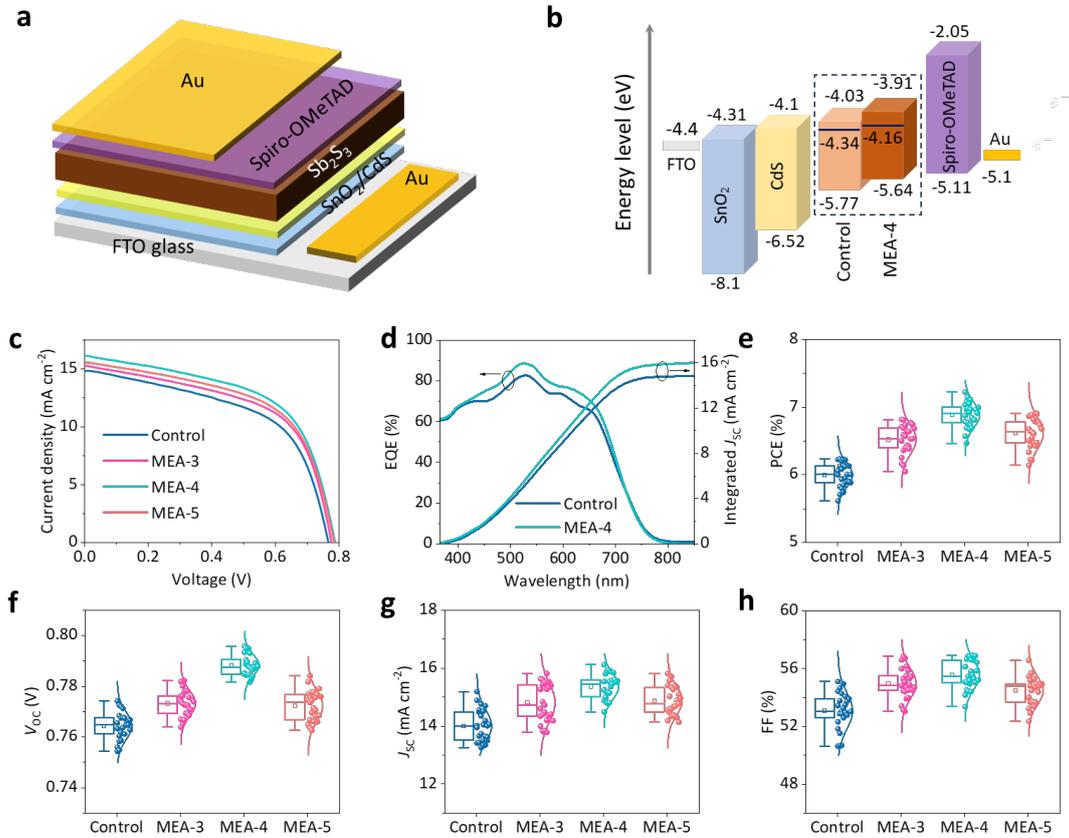

**Fig. 3 Photovoltaic performances of $Sb_2S_3$ solar cells under AM1.5G illumination. a** Schematic illustration of the device configuration of planar $Sb_2S_3$ solar cells. **b** Band alignment of the components of $Sb_2S_3$ solar cells. **c** *J-V* curves of best-performing control and MEA-$Sb_2S_3$ solar cells, measured under 1-sun (AM1.5G, 100 mW cm$^{-2}$) illumination. **d** EQE spectra of the control and MEA-4 $Sb_2S_3$ solar cells. **e-h** Statistics of the performance parameters of the control and MEA-$Sb_2S_3$ devices obtained with the addition of different concentration of MEA.

**Table 1** Photovoltaic performance parameters of the control and MEA-$Sb_2S_3$ solar cells with the addition of different concentration of MEA, measured under 1-sun (AM1.5G, 100 mW cm$^{-2}$) illumination. Format: mean ± standard deviation (best value)

| Devices | $V_{OC}$ (mV) | $J_{SC}$ (mA cm$^{-2}$) | FF (%) | PCE (%) |
| --- | --- | --- | --- | --- |
| Control | 764 ± 5 (774) | 14.01 ± 0.55 (15.19) | 53.07 ± 1.27 (55.10) | 5.99 ± 0.18 (6.22) |
| MEA-3 | 773 ± 4 (782) | 14.81 ± 0.66 (15.82) | 54.96 ± 1.04 (56.85) | 6.52 ± 0.21 (6.81) |
| MEA-4 | 788 ± 4 (795) | 15.36 ± 0.43 (16.12) | 55.57 ± 0.98 (56.92) | 6.88 ± 0.18 (7.22) |
| MEA-5 | 772 ± 5 (784) | 14.87 ± 0.49 (15.81) | 54.48 ± 0.96 (56.58) | 6.61 ± 0.23 (6.91) |

It is worth noting that, since the addition of MEA affects the thickness of the $Sb_2S_3$ films, the control and MEA-$Sb_2S_3$ devices were constructed based on absorber films with different thicknesses. In order to reveal the impact of the film thickness on the device performance, we further compared the performance of $Sb_2S_3$ solar cells based on the $Sb_2S_3$ films with different thickness prepared by using conventional hydrothermal precursor solutions without the addition of MEA. It is found that, the device based on the 430 nm-thick $Sb_2S_3$ film that shares a similar thickness as that of the MEA-4 film involves a slight efficiency drop compared to the 339 nm-thick film (Supplementary Fig. 11a and b). This indicates that the performance enhancement of the MEA-$Sb_2S_3$ devices is not caused by an increase in light absorption. The SEM image (Supplementary Fig. 11c) further reveals that, compared to the 339 nm-thick film, the grain size of the 430 nm-thick $Sb_2S_3$ films shows no negligible changes. However, the XRD patterns reflect that the (hk1) orientation of this 430 nm-thick $Sb_2S_3$ film is also slightly enhanced compared to the 339 nm-thick film (Supplementary Fig. 11d), similar to the scenario for MEA-$Sb_2S_3$ films. Therefore, it is reasonable to speculate that the enhanced (hk1) orientation of the MEA-4 $Sb_2S_3$ film might be related to the increase of film thickness. The underlying mechanism for the impact of the film thickness on the film orientation needs further study. In addition, a slight drop in the device efficiency was observed when the concentration of MEA increased up to 5 μg/mL, and this might be caused by a continuous increase in the film thickness. To verify this hypothesis, we optimized the thickness of absorption layer based on the device performance of $Sb_2S_3$ solar cells (Supplementary Fig. 12a and b and Supplementary Table 2). As reflected by the corresponding thickness of the MEA-4 films prepared for different times of hydrothermal growth (Supplementary Fig. 12c-e), the optimal thickness for $Sb_2S_3$ layer is around 428 nm, which can be obtained after 180 min of hydrothermal deposition. Further increases in the film thickness would result in the performance degradation of $Sb_2S_3$ solar cells. Based on such discussions, we conclude that the positive impact of the addition of MEA into the precursor solutions should be responsible for the enhanced performance of MEA-$Sb_2S_3$ solar cells. Moreover, in Section 2.1, we propose that -OH and -$NH_2$ groups in MEA contribute to form coordination complexes with $Sb^{3+}$ in the precursor solution. In order to confirm this, we also investigated the use of diethanolamine (DEA) and triethanolamine (TEA), which share similar coordination groups, as the additives in the precursor solution to deposit $Sb_2S_3$ films (Supplementary Fig. 13). The corresponding device performance reveals that the incorporation of DEA and TEA additives also contribute to efficiency improvements of $Sb_2S_3$ solar cells (Supplementary Fig. 14).

We further performed various characterization of $Sb_2S_3$ solar cells to study the device physics and explore the underlying mechanisms responsible for the performance enhancements. Space charge limited current (SCLC) density measurements of electron-only device (FTO/CdS/$Sb_2S_3$/PCBM/Au) are given in Fig. 4a. Generally, the curves can be divided into three parts, the Ohmic region (low voltage), the trap-filled limiting (TFL) region (intermediate voltage), and the trap-free region (high voltage). At the low voltage region, the curve is generally linear. When the voltage exceeds the trap-filled limiting voltage ($V_{TFL}$), the current increases dramatically, implying that the injected carriers have completely occupied the trap states[26]. The $V_{TFL}$ values for the control and MEA-4 device are estimated to be 0.486 V and 0.414 V, respectively. The trap state density ($n_\tau$) of $Sb_2S_3$ films can be evaluated according to the equation $V_{TFL} = qn_\tau L^2 / 2\varepsilon_r\varepsilon_0$, where $q$ is the elementary charge, $n_\tau$ is the trap state density, $L$ is the thickness of the absorber film (here 339 nm and 427 nm for the control and MEA-4 $Sb_2S_3$ films, respectively), $\varepsilon_0$ is the permittivity of free space ($8.85 \times 10^{-12}$ F m$^{-1}$), and $\varepsilon_r$ represents the relative permittivity of Sb$_2$S$_3$ (6.67). Then the estimated $n_\tau$ for the control device and the MEA-4 Sb$_2$S$_3$ device are $3.12 \times 10^{15}$ and $1.67 \times 10^{15}$ cm$^{-3}$, respectively. The reduced trap-state density for the MEA-4 sample indicates that the addition of MEA can effectively reduce the density of trap states in Sb$_2$S$_3$ films[36]. We also performed the dependence of $J_{SC}$ and $V_{OC}$ on the light intensity for both devices to reveal the trap-assisted charge recombination loss mechanisms, as shown in Fig. 4b and c. The relationship between $J_{SC}$ (or $V_{OC}$) and the light intensity $I$ can be described by the equations $J_{SC} \propto I^\alpha$ and $V_{OC} \propto (nk_B T/q)\ln(I)$, where $I$ is the light intensity, $k_B$ is the Boltzmann constant, $T$ is the absolute temperature, $q$ is the elementary charge, $\alpha$ and $n$ reflect the level of charge-carrier recombination[37,38]. The calculated $\alpha$ for the control and MEA-4 Sb$_2$S$_3$ device are 0.904 and 0.941, respectively. Ideally $\alpha$ equals 1, and $\alpha <$ 1 means the loss of photogenerated carriers caused by the incomplete charge collection. The increment in $\alpha$ for the MEA-4 device confirms the improvement of the carrier collection rate[38,39]. For trap-free solar cells, the slope of $V_{OC}$ versus ln($I$) is $k_B T/q$. By performing linear fitting of these curves, the slopes for the control and MEA-4 devices are obtained to be 1.53$k_B T/q$ and 1.41$k_B T/q$, respectively. The reduced slope for the MEA-4 device indicates that the trap-assisted charge-carrier recombination is suppressed. Since the Urbach energy ($E_U$) is capable of evaluating the band tailing effect, we derived $E_U$ value from the EQE spectrum by performing linear fitting of the ln(EQE) vs $E_U$ plots (Supplementary Fig. 15)[20]. The $E_U$ values of the control and MEA-4 device were fitted to be 32.36 and 27.98 meV, respectively. The reduction in $E_U$ suggests the suppression of nonradiative recombination probably due to the decrease of defect density, thus explaining the improvement of $V_{OC}$ for MEA-Sb$_2$S$_3$ devices.

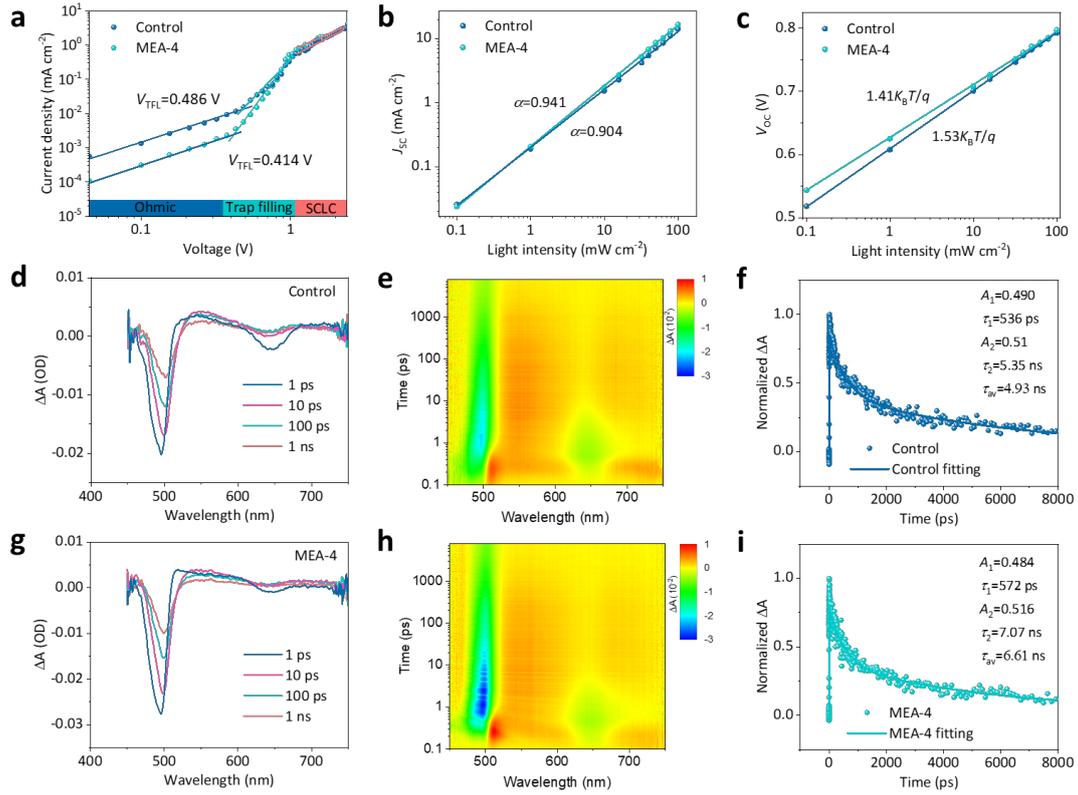

**Fig. 4 Device physics and charge-carrier kinetics. a** SCLC measurements of the control and MEA-4 devices based on the electron-only device configuration of FTO/CdS/Sb$_2$S$_3$/PCBM/Au. **b, c** The dependence of $J_{SC}$ and $V_{OC}$ on the light intensity for the control and MEA-4 Sb$_2$S$_3$ solar cells. **d, g** Time-resolved absorption spectra obtained at 1, 10, 100 and 1000 ps, excited by a 400 nm laser pulse, **e, h** 2D TA spectroscopy pseudo-color images for the time-resolved absorption spectra, and **f, i** Transient kinetic decay (scatter) and corresponding bi-exponential curve fittings (solid line) monitored at 560 nm of the control and MEA-4 Sb$_2$S$_3$ films. $\Delta A$ is defined as the variation of absorption.

The *J-V* curves of the control and MEA-4 solar cells measured under dark conditions are collected to evaluate the carrier recombination and collection in the devices (Supplementary Fig. 16a). According to the abrupt junction *J-V* equation as expressed in Equation (S15), and its formula manipulation as given in Equations (S16-S18), the parameters of diode ideality factor (*A*), the reverse saturation current density ($J_0$), the series resistance ($R_S$), and the shunt conductance (*G*, i.e., $1/R_{SH}$) can be extracted (Supplementary Fig. 16b-d and Supplementary Table 3)[40]. The detailed analysis is given in Supplementary Note S3. It can be seen that, compared to the control device, the MEA-4 device has lower $R_S$ and $G$, which implies an improvement in the carrier extraction capability. The $A$ values extracted from the control and MEA-4 devices are 2.33 and 1.87, respectively; the reduced $A$ value indicates the suppression of defect states in the absorber films[25]. The calculated $J_0$ is reduced from $9.43\times10^{-6}$ mA cm$^{-2}$ for the control device to $1.49\times10^{-6}$ mA cm$^{-2}$ for the MEA-4 device. The main reason for the generation of $J_0$ is the charge-carrier recombination caused by deep-level defects in the devices[41], and the reduced $J_0$ implies the suppression of the defect-induced recombination, contributing to the increase of $V_{OC}$ for the MEA-4 device.

The electrochemical impedance spectroscopy was also employed to study the carrier recombination and transport behaviors of devices. The Nyquist plots of the impedance spectra for the control and MEA-4 devices were measured under dark conditions at the bias voltage of 0.7 V (Supplementary Fig. 17), and the resistance and capacitance parameters were obtained by performing the curve fitting based on the equivalent circuit diagram given in the inset (Supplementary Table 4). Here $R_{ser}$, and $R_{rec}$ in the equivalent circuit represent the internal series resistance and recombination resistance, respectively. CPE consists of the interface capacitance (CPE-T) and the ideal capacitance (CPE-P)[42]. $R_{rec}$ is related to charge-carrier recombination and its value is equal to the diameter of the semicircle in the Nyquist plots. As shown, the MEA-4 device delivers an increased recombination resistance at the CdS/Sb$_2$S$_3$ interface ($R_{rec}$=16.02 kΩ cm$^2$) compared to the control device ($R_{rec}$=9.62 kΩ cm$^2$). The significant increase in the recombination resistance is suggested to suppress the charge recombination and improve the charge collection in solar cells.

Ultrafast transient absorption (TA) spectroscopy measurements were performed to understand the charge-carrier kinetics of FTO/SnO$_2$/CdS/Sb$_2$S$_3$ films. As shown in Fig. 4d and g, the TA spectra of both the control sample and the MEA-4 sample display distinct negative ground state bleach (GSB) peaks and positive photoinduced absorption (PIA) peaks. The GSB peaks observed at the wavelengths of 460-510 nm and 610-680 nm can be attributed to the state filling of CdS and the ground state absorption of Sb$_2$S$_3$, respectively. The PIA peak at the wavelength of 520 to 620 nm can be ascribed to the formation of sulfide radicals (S$^-$) as a result of the localization of photogenerated holes on the S atom within the $Sb_2S_3$ lattice[43,44]. The transient dynamics are extracted from the pseudocolor images for the TA spectra of both samples (Fig. 4e and h). The transient kinetic decays monitored at 560 nm for both the control and MEA-4 films are presented in Fig. 4f and i, respectively, which can be well fitted by a phenomenological biexponential equation $y = A_1 \exp(-t/\tau_1) + A_2 \exp(-t/\tau_2)$, where $\tau_1$ and $\tau_2$ denote short-lived and long-lived carrier lifetimes. The fitting results are listed in Supplementary Table 5. The gradual decrease of the PIA peak can be attributed to the decay of trapped holes, *i.e.*, the $S^-$ species, which we here attribute to the nonradiative carrier recombination in $Sb_2S_3$ films. The MEA-4 sample delivers higher $\tau_{av}$ values (6.61 ns) compared to the control sample (4.93 ns). The longer lifetime in the MEA-4 sample suggests the suppression of the bulk charge-carrier recombination. Therefore, the TA analysis indicates a slower charge recombination rate in MEA-$Sb_2S_3$ films. The extended lifetime of minority hole carriers would contribute to the enhancement in $V_{OC}$[21].

## 2.4. IPV performance of $Sb_2S_3$ devices and demonstration for powering IoT electronics

The bandgaps of the $Sb_2S_3$ films obtained by hydrothermal deposition in this study are 1.73-1.74 eV, which are very close to the optimum value of bandgap for indoor photovoltaics under white light illumination (1.9-2.0 eV)[7], as shown in Fig. 5a. Fig. 5b shows the emission spectra of AM 1.5G, 3000K WLED and 2700K FL, which reveals that the intensity of indoor light sources is 100-1000 times lower than that of 1-sun[8]. Firstly, we measured the emission power spectra of 3000 K WLED and 2700 K FL light sources at 200, 500, and 1000 lux illuminance, as shown in Fig. 5c and Supplementary Fig. 18. The corresponding optical power densities of 3000K WLED were calculated to be 291.9 µW cm$^{-2}$ (1000 lux), 145.2 µW cm$^{-2}$ (500 lux), 62.0 µW cm$^{-2}$ (200 lux), respectively, while the 2700 K FL delivers 296.7 µW cm$^{-2}$ (1000 lux), 153.1 µW cm$^{-2}$ (500 lux), 64.6 µW cm$^{-2}$ (200 lux), respectively (Supplementary Table 6). All IPV measurements were performed in a black box to prevent the influence of stray light from the environment (Supplementary Fig. 19)[17]. The *J-V* curves for the best-performing MEA-4 $Sb_2S_3$ solar cell measured under WLED and FL illumination are shown in Fig. 5d and e, respectively. The corresponding IPV performance parameters are summarized in Table 2. As shown, under the WLED illuminations of 1000, 500, and 200 lux, the PCEs of this device are 17.55%, 17.17%, and 16.41%, respectively, with the output power densities being 51.23, 24.93, 10.18 μW cm$^{-2}$, respectively; under the FL illuminations of 1000, 500, and 200 lux, the device delivers the PCEs of 16.54%, 15.92%, and 15.51%, respectively, with the output power densities being 49.09, 24.37, 10.02 μW cm$^{-2}$, respectively. As expected, $Sb_2S_3$ solar cells exhibit excellent IPV performance under low-intensity indoor light illuminations. To the best of our knowledge, the IPV efficiency of 17.55% is the highest reported thus far for $Sb_2S_3$ photovoltaics under indoor lighting. The dependences of photovoltaic parameters (PCE, $V_{OC}$, $J_{SC}$, FF, $R_S$, and $R_{SH}$) of the MEA-4 device on the indoor light intensity are depicted in Supplementary Fig. 20. As shown, $J_{SC}$ is linearly proportional to the illumination intensity, $V_{OC}$ decreases approximately logarithmically with the illumination intensity, and FF involves very small changes under such low illumination intensities. This enables the $Sb_2S_3$ device to yield the best IPV efficiency under 1000 lux. Moreover, compared to the device performance measured under the standard AM1.5G illumination, the low-intensity illumination leads to a significant reduction in $J_{SC}$ and $V_{OC}$ and a remarkable enhancement in FF under the indoor light illumination[45]. Such photovoltaic parameters involve similar variation trend with those previously reported in the literature[18]. The detailed discussions are given in Supplementary Note S4. We further recorded the device stability of the unencapsulated $Sb_2S_3$ solar cell stored in a cabinet with 15% relative humidity at ambient temperature (Fig. 5f). The device retains 96.75% of its initial efficiency after one-month storage. The excellent device stability under indoor environments demonstrates the great prospect of $Sb_2S_3$ solar cells for IoT applications.

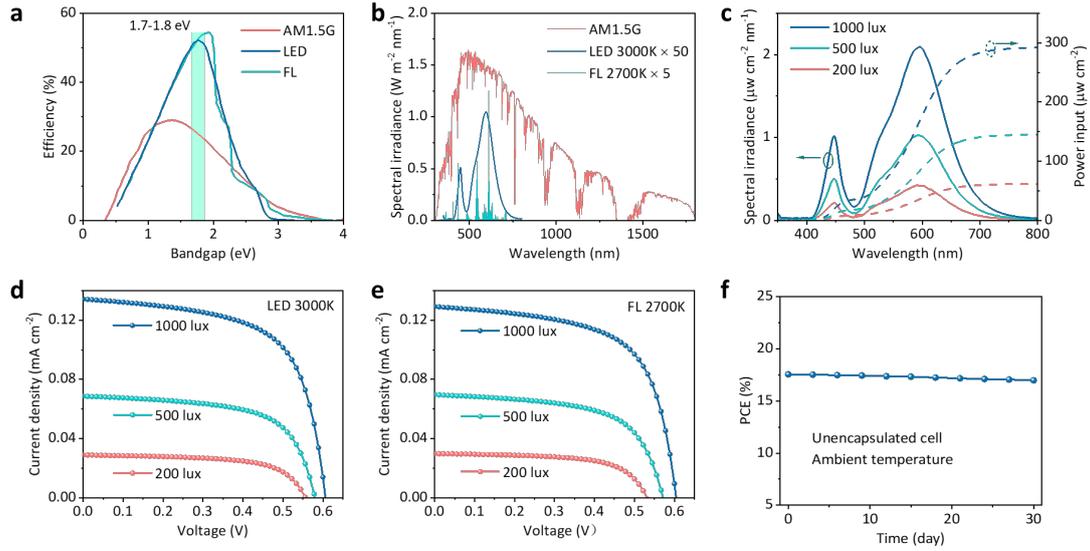

**Fig. 5 IPV performance of Sb$_2$S$_3$ solar cells. a** Calculated efficiency limit of Sb$_2$S$_3$ devices under 1-sun (AM1.5G, 100 mW cm$^{-2}$), 3000 K WLED at 1000 lux, and 2700 K FL at 1000 lux illuminations, respectively.[46] **b** Comparison of the emission spectra of AM1.5G solar, 3000 K WLED at 1000 lux, and 2700 K FL at 1000 lux. The spectral intensities of WLED and FL are amplified by 50 and 5 times, respectively, in order for these spectra to be legible in this plot. **c** Spectral irradiance and integrated power density of a 3000 K WLED at 1000, 500, and 200 lux. **d, e** $J$–$V$ curves of the MEA-4 Sb$_2$S$_3$ solar cells measured under the illumination of LED and FL at 1000, 500, and 200 lux. **f** Device stability of unencapsulated MEA-4 Sb$_2$S$_3$ solar cells stored in a cabinet with 15% relative humidity at ambient temperature.

**Table 2** Indoor photovoltaic performance of MEA-4 Sb$_2$S$_3$ solar cells measured under WLED (3000 K) and white light FL (2700 K) at 1000, 500, and 200 lux.

| MEA-4 device | $V_{OC}$ (mV) | $J_{SC}$ (mA cm$^{-2}$) | FF (%) | PCE (%) | $R_S$ (kΩ cm$^2$) | $R_{SH}$ (kΩ cm$^2$) |
|---|---|---|---|---|---|---|
| 1-sun | 787 | 16.12 | 56.92 | 7.22 | 0.007 | 0.23 |
| 1000 lux LED | 606 | 0.134 | 63.09 | 17.55 | 0.44 | 48.31 |
| 500 lux LED | 579 | 0.0685 | 62.86 | 17.17 | 0.76 | 86.57 |
| 200 lux LED | 557 | 0.0289 | 63.22 | 16.41 | 2.17 | 206.66 |
| 1000 lux FL | 604 | 0.129 | 63.00 | 16.54 | 0.44 | 49.47 |
| 500 lux FL | 571 | 0.0696 | 61.32 | 15.92 | 0.90 | 75.55 |
| 200 lux FL | 531 | 0.0297 | 63.54 | 15.51 | 1.97 | 226.47 |

Having developed efficient $Sb_2S_3$ IPVs, we are now in the position to prototype these for the first time for powering IoT wireless electronics. It is known that a large number of IoT devices typically require very little energy with power consumption ranging from several μW to a few mW, which depends on the application and communication protocol[9]. In this work, we used the RSL 10 solar cell multi-sensor platform (RSL10-SOLARSENS-GEVK) manufactured by ON Semiconductor, which is a comprehensive development platform for battery-less IoT applications in smart buildings, smart homes, and industrial sectors (Supplementary Fig. 21). This board is based on the industry's lowest-power Bluetooth® low-power radio (RSL10) and features sensors for environmental sensing: the BME280 smart environmental sensor. The platform also features a lightweight, low-profile 47 F storage capacitor, a programming and debugging interface, and an interface to a solar cell. To drive this sensor platform via an external power supply, a solar module with an $V_{OC}$ higher than 2.65 V and an average input power exceeding 50 μW is needed. The $J$-$V$ curve of as-prepared 1 cm$^2$ area MEA-4 $Sb_2S_3$ solar cell (Supplementary Fig. 22) demonstrates a PCE of 15.77% under 1000 lux WLED illumination, coupled with a $V_{OC}$ of 0.596 V, a $J_{SC}$ of 124 μA cm$^{-2}$ and an FF of 62.39%. Compared with the 0.06 cm$^2$ small-area cell, the device performance of this 1 cm$^2$ area device only shows about a 10% decrease in PCE. We then obtained an $Sb_2S_3$ module (5×1 cm$^2$) by connecting five $Sb_2S_3$ devices together in series. As shown in Fig. 6a, the $J$-$V$ curve of this $Sb_2S_3$ module has a $V_{OC}$ of 2.92 V and an output power of 187.17 μW at maximum power point under 1000 lux WLED illumination, which is sufficient to drive the RSL10-SOLARSENS-GEVK platform. Compared with commercial a-Si:H IPV products in the market, the $Sb_2S_3$ module here is of competitive performance in terms of PCE, $V_{OC}$ and output power[3,13]. The connection between the sensor and the $Sb_2S_3$ module is shown in Fig. 6b. Fig. 6c illustrates an IPV-driven sensor device, in which the $Sb_2S_3$ module converts light energy into electrical energy and drives the sensor to operate stably. The sensor communicates with the mobile phone via Bluetooth Low Energy, which enables the real-time monitoring and collection of environmental data. Supplementary Fig. 23 presents plots of the temperature, humidity, and atmospheric pressure measured in the laboratory by the $Sb_2S_3$ module-driven sensor platform running continuously under 1000 lux WLED illumination for 140 min in ambient air without encapsulation. Furthermore, we placed it on a desk in a common office area (with a light intensity of ~726 lux) to successfully monitor the environmental parameters for nearly 14 hours, as shown in Fig. 6d-f, confirming that the $Sb_2S_3$ module can drive IoT devices under indoor light continuously and stably.

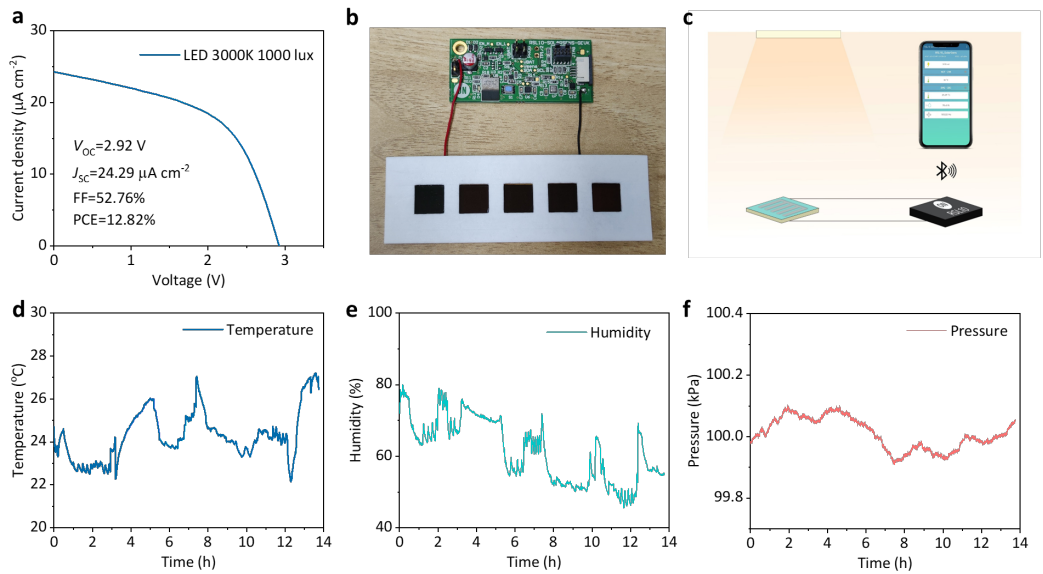

**Fig. 6 Demonstration for powering IoT electronics. a** *J-V* curve of $Sb_2S_3$ solar module (5×1 cm$^2$). **b** The picture showing the connection between the sensor platform and the $Sb_2S_3$ module. **c** Schematic circuit diagram of an IPV-driven sensor device. **d-f** Temperature, humidity, and atmospheric pressure variations recorded by the IPV-driven sensor in the office for nearly 14 hours.

## 3. Conclusion

In this work, we demonstrate an effective additive engineering strategy to improve the optoelectronic properties of $Sb_2S_3$ films and enhance the photovoltaic performance of corresponding planar $Sb_2S_3$ solar cells. The multifaceted role of MEA plays a pivotal part in regulating the nucleation and growth of $Sb_2S_3$ films for enhanced heterogeneous nucleation on the substrate. Compared to the control sample, the MEA-modulated $Sb_2S_3$ films show reduced GB density, optimized band positions, and increased carrier concentration, affording suppressed charge-carrier recombination and efficient charge collection in solar devices. The corresponding MEA-$Sb_2S_3$ solar cell achieves a state-of-the-art efficiency of 7.22% under AM1.5G illumination, much higher than that of the control device (6.22%). This device delivers an IPV efficiency of 17.55% under 1000 lux WLED illumination, which, to our knowledge, is the highest reported so far for $Sb_2S_3$ solar cells. Furthermore, we constructed an $Sb_2S_3$ module (5×1 $cm^2$) and successfully reported the use of this module to power IoT wireless devices for the first time. This study demonstrates the broad prospect of $Sb_2S_3$ solar cells for IPV applications.

**Acknowledgements**

X.C. and X.S. contributed equally to this work. The financial support by the National Natural Science Foundation of China (nos. 52371219 and U19A2089), and the National Natural Science Foundation of Hefei City (no. 2022024) is greatly acknowledged. Y. F., J. Y. and R. L. Z. H. acknowledge funding from a UKRI Frontier Grant (no. EP/X022900//1). Y.-T. H. and R. L. Z. H. acknowledge funding from the Engineering and Physical Sciences Research Council (grant no. EP/V014498/2). R. L. Z. H. thanks the Royal Academy of Engineering for support through the research fellowships scheme (no. RF/201718/17101), and also the Henry Royce Institute for support through the Industrial Collaborative Programme, funded by EPSRC (no. EP/X527257/1).

**Author contributions**

R.Z. and X.C. designed the experiments and analyzed the data, and also conceived of the idea for this manuscript, with input from R.L.Z.H on the idea of IPV development and prototyping. X.C. carried out the experiments and device optimizations. X.S. carried out the DFT calculations. J.Z., P.X., L.W., Y.F., J.Y., Y.-T. H., B.Y., D.-J. X, T.C.,

J.C., R.L.Z.H. and R.Z. assisted in experiments and data analysis. R.Z., X.C., and R.L.Z.H. wrote the manuscript. All authors commented on the manuscript.

**Conflict of interest**

The authors declare no conflict of interest.